\documentclass{iaus}
\usepackage{graphicx}

\title[~~Variation in the dust spectral index across M33] 
{Variation in the dust spectral index across M33}

\author[Tabatabaei et al.]   
{Fatemeh S. Tabatabaei$^1$,
Jonathan Braine$^2$,
Carsten Kramer$^3 $,
Manolis Xilouris$^4$,
Mederic Boquien$^5$,
Simon Verley$^6$,\\
Eva Schinnerer$^7$,
Daniela Calzetti$^8$,
Francoise Combes$^9$,
Frank Israel$^{10}$,
Christian Henkel$^{11}$
\and The HerM33es Team}

\affiliation{$^1$Max-Planck-Institut f\"ur Astronomie,  K\"onigsthul 17,
69117-Heidelberg, Germany \\ email: {\tt taba@mpia.de} \\[\affilskip]
$^2$Laboratoire d'Astrophysique de Bordeaux, Universit\'e de Bordeaux, 33271 Floirac, France \\email:{\tt braine@obs.u-bordeaux1.fr} \\[\affilskip]
$^3$Instituto Radioastronomia Milimetrica, 18012 Granada, Spain\\email:{\tt kramer@iram.es}\\[\affilskip]
$^4$Institute of Astronomy and Astrophysics, National Observatory of Athens, 
GR-15236 Athens, Greece, email:{\tt xilouris@astro.noa.gr}\\[\affilskip]
$^5$ Laboratoire d'Astrophysique de Marseille, UMR 6110 CNRS, Marseille, France
 \\email:{\tt mederic.boquien@oamp.fr}\\[\affilskip]
$^6$Department F\'isica Te\'orica y del Cosmos, Universidad de Granada, Spain 
\\email:{\tt simon@ugr.es}\\[\affilskip]
$^7$Max-Planck-Institut f\"ur Astronomie, K\"onigsthul 17,
69117-Heidelberg, GermanyGermany\\ email: {\tt schinner@mpia.de}\\[\affilskip]
$^8$Department of Astronomy, University of Massachusetts, Amherst, MA 01003, USA \\email:{\tt calzetti@astro.umass.edu}\\[\affilskip]
$^9$ Observatoire de Paris, 75014 Paris, France \\ email: {\tt francoise.combes@obspm.fr}\\[\affilskip]
$^{10}$Sterrewacht Leiden, Leiden University, PO Box 9513, 2300 RA, Leiden, The Netherlands \\ email: {\tt israel@strw.leidenuniv.nl}\\[\affilskip]
$^{11}$Max-Planck-Institut f\"ur Radioastronomie,  Auf dem H\"ugel 69,
53121-Bonn, Germany\\ email: {\tt chenkel@mpifr.de}}

\pubyear{2011} 
\volume{284}  
\pagerange{1--12}
\jname{The Spectral Energy Distribution of Galaxies}
\editors{R.J. Tuffs \&  C.C.Popescu, eds.}
\begin{document}

\maketitle

\begin{abstract}
Using the Herschel PACS and SPIRE FIR/submm data, we investigate variations in the dust spectral index  $\beta$ in the nearby spiral galaxy M33 at a linear resolution of 160\,pc.  We use an iteration method in two different approaches, single and two-component modified black body models. In both approaches, $\beta$ is higher in the central disk than in the outer disk similar to the dust temperature. There is a positive correlation between $\beta$ and H$\alpha$ as well as with the molecular gas traced by CO(2-1). A Monte-Carlo simulation shows that the physical parameters are better constrained when using the two-component model. 

\end{abstract}

\firstsection 
\section{Introduction}

Inferring  the nature of dust and  dust heating requires  knowledge  about  the dust emissivity, and its power law index. There is some  evidence  supporting  variation of  dust  emissivity with environmental  conditions  (e.g. Lisenfeld et al. 2000, Paradis et al. 2009).  The dust  spectral  index  could  change  depending on grain properties,  e.g.  geometry, size  distribution, and  chemical  composition.  Dust grains might be affected  by  different   physical  processes  like  shattering,  sputtering,  grain - grain  collision, condensation  and   coagulation.   Hence,  depending on  recent  star formation  activities, dust/gas content and metallicity changes across galaxies, variations in the dust spectral index in inner/outer disks  or 
arm/inter-arm regions may occur.   Herschel  observations of the nearest Scd galaxy M33, (HerM33es, Kramer et al. 2010, Boquien et al. 2010)  ideally enable  us  to  derive  distribution of $\beta$ across M33.
 
We use the Herschel PACS \& SPIRE data at 160, 250, 350, and  500$\mu$m   together with  the  Spitzer MIPS  data  at 70$\mu$m  to investigate  the  physical properties of  a standard dust model (Big grains+vsg+PAH, Kr\"ugel 2003 and references therein). These wavelengths spread over a range where the dust emission mainly emerges from the big grains, which are believed to be in thermal equilibrium with the interstellar radiation field. 

\begin{figure}
\begin{center}
\resizebox{10cm}{!}{\includegraphics*{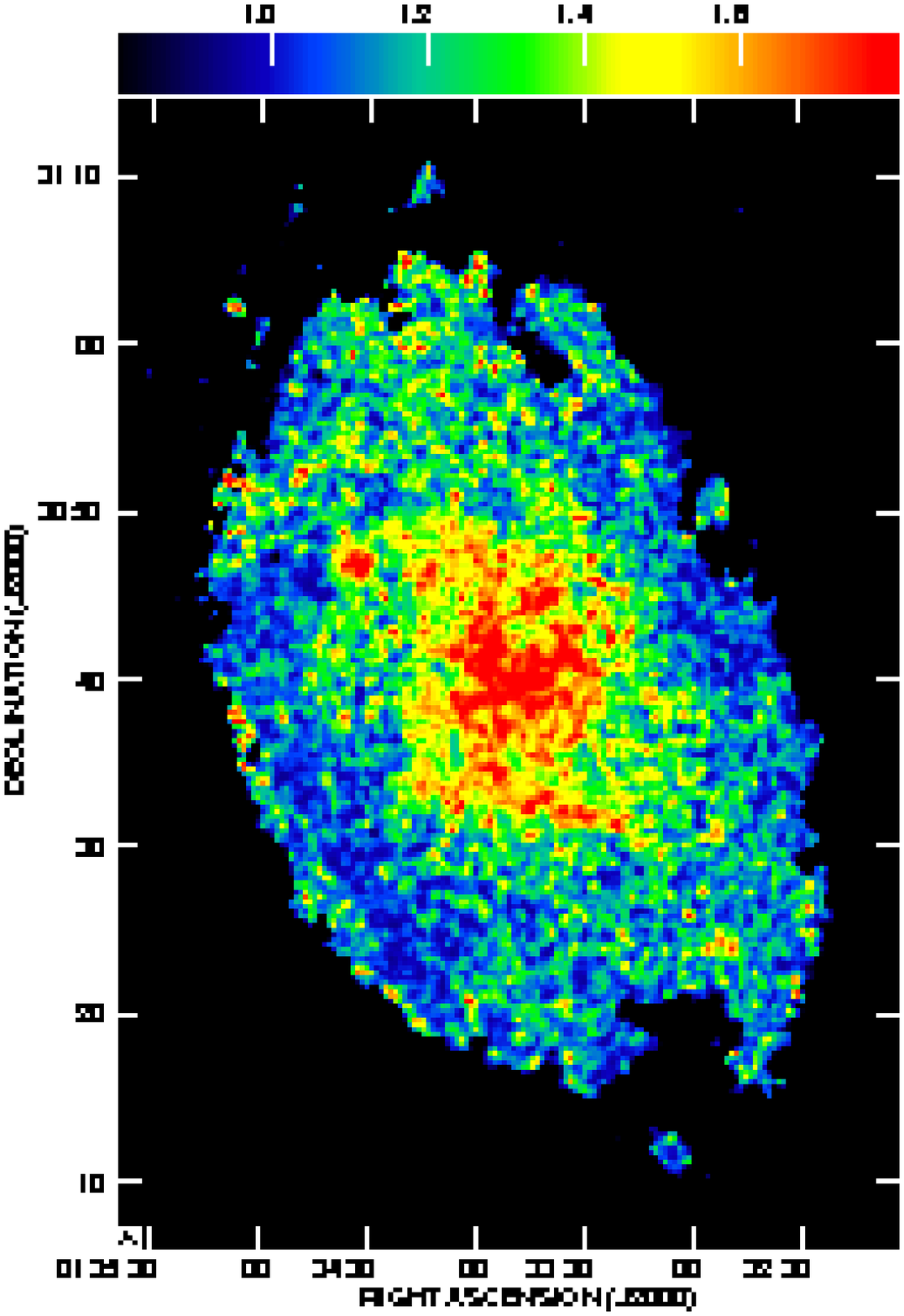}\includegraphics*{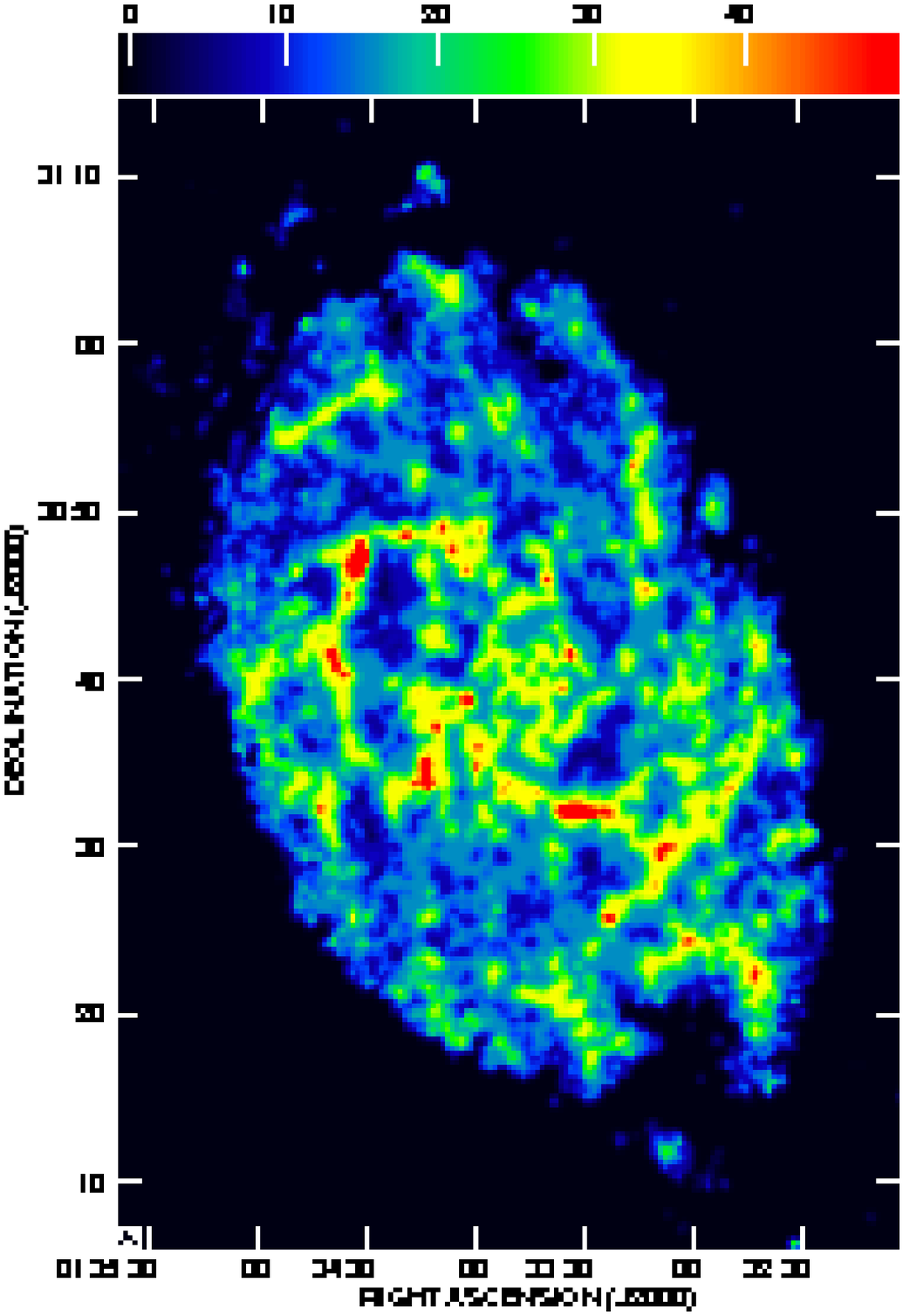}}
\caption[]{Distribution of dust spectral index (right) and column density (in $\mu$g\,cm$^{-2}$, right)  across M33, based on the single-component approach.   }
\end{center}
\end{figure}

\section{Analysis and Results}
We  use  the Newton-Raphson iteration method to derive $T$, $\beta$ and column density, pixel by pixel, in two different approaches:\\
1) We assume that dust emitting at the SPIRE bands is mainly heated by  the  interstellar  radiation field  (ISRF) and obeys  single - component gray body radiation. In this case we use the data at longer wavelengths (250, 350 and 500$\mu$m).  \\
2) We  consider another dust component  heated  by young  massive  stars with higher temperature  (warm dust) besides the dust heated  by the ISRF (cold dust), hence, dust  emits as a two-component gray body. Here, we use the data from 70$\mu$m to 500$\mu$m wavelengths.\\
Using a Monte-Carlo simulation, we further investigate the reliability of each approach and estimate the errors and uncertainties.
We derive the following results:  \\
\\
\underline{\large Single Component Approach}
As shown in Fig.~1, in the central disk ($R<4$ kpc), the dust spectral index is higher in the central disk ($\beta >$1.4) than in the outer disk. Similarly, both T and column density are higher in the central disk. The flocculent spiral structure of M33 is evident in the column density map. The average T and $\beta$ for the entire galaxy are 22$\pm$9 K  and 1.3$\pm$0.5, respectively.  
\\
\underline{\large Two Component Approach}
The cold and warm dust components are separated assuming a variable $\beta$ for the dominant colder component. Although  we have to use a fixed value of $\beta_{\rm warm}$ for the warm dust, we repeat the analysis for different possibilities of $\beta_{\rm warm}$=1, 1.5, and 2.  Interestingly, we find that $\beta$ is again higher in the central disk and even shows  the arm/inter-arm contrast, irrespective of $\beta_{\rm warm}$ assumed.  Also in all three cases, the  cold dust  follows  spiral  arms and dust lanes  and  is denser where the temperature is lower. The warm dust (both temperature and column density)  follows the distribution of H$\alpha$ and star forming regions.  For the entire galaxy, we derive T$_{\rm cold}$=\,16 $\pm$\,3\,K, T$_{\rm warm}$=\,40\,$\pm$\,13\,K, and $\beta$=\,1.7\,$\pm$\,0.4 assuming $\beta_{\rm warm}$=2. Our simulations show that the two-component approach, generally, constrains the parameters better than the single-component approach, particularly the temperature.

\section{Correlation with ISM Tracers}
The M33's outer/inner disk and even the spiral arms are visible in the $\beta$-maps. 
Having similar distribution as of the molecular gas and radiation field traced by H$\alpha$, $\beta$ seems to have a meaningful change across the galaxy.  
Figure 2 shows correlations  between $\beta$ and  H$\alpha$, CO(2-1) and HI integrated intensities (Gratier et al. 2010) as tracers of the ionized, molecular, and atomic phases of the ISM, respectively.  In both approaches, $\beta$ is found to be better correlated with both H$\alpha$ and CO than with HI.  This could indicate that $\beta$ is not the same in star forming and non-starforming regions in a galaxy. Star formation could in principle influence the grain composition (via a change in metallicity) and size distribution (via shattering of dust grains due to strong shocks), which could modify the dust emissivity index i.e.  $\beta$. 



\begin{figure}
\begin{center}
\resizebox{\hsize}{!}{ \includegraphics*{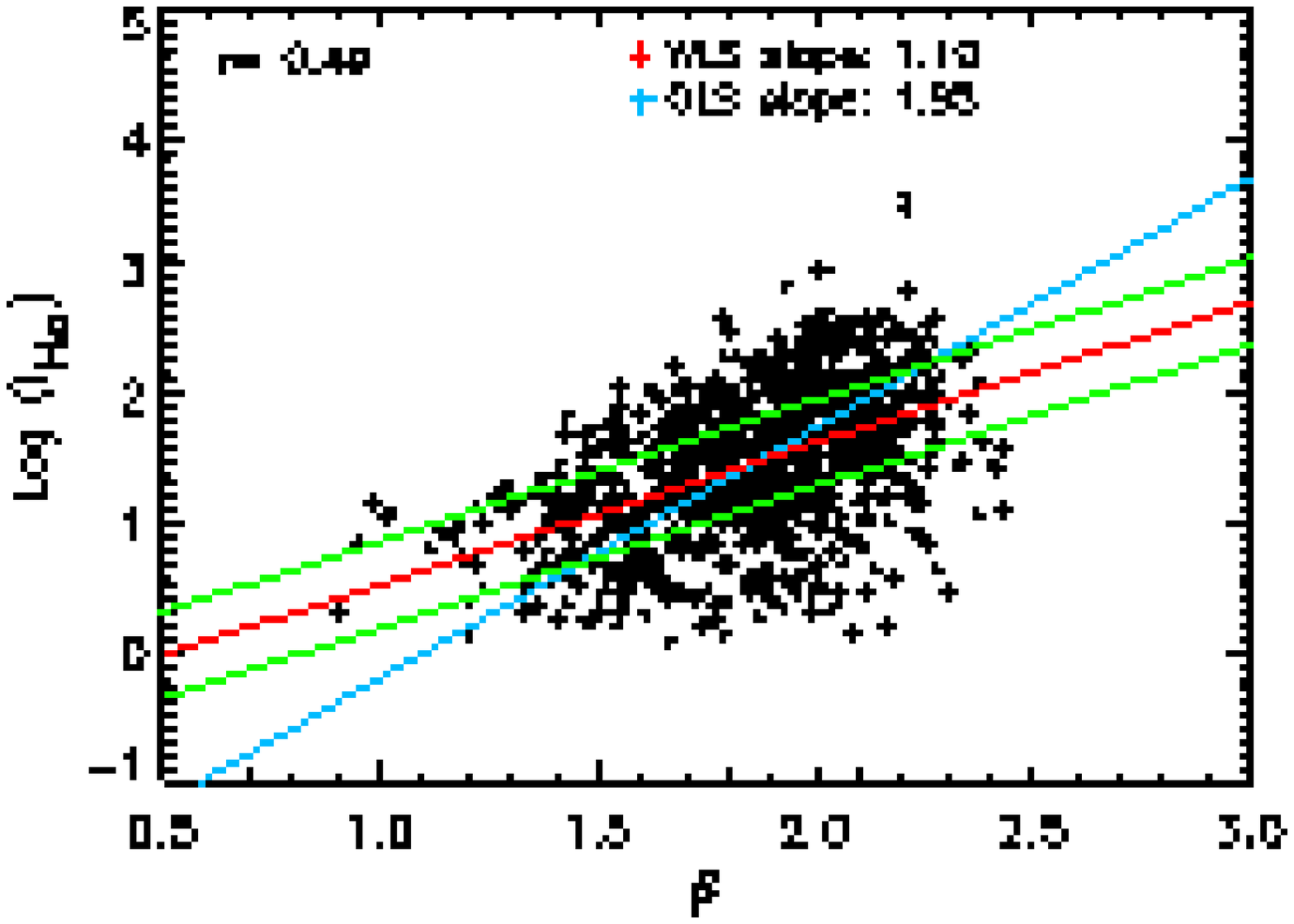}\includegraphics*{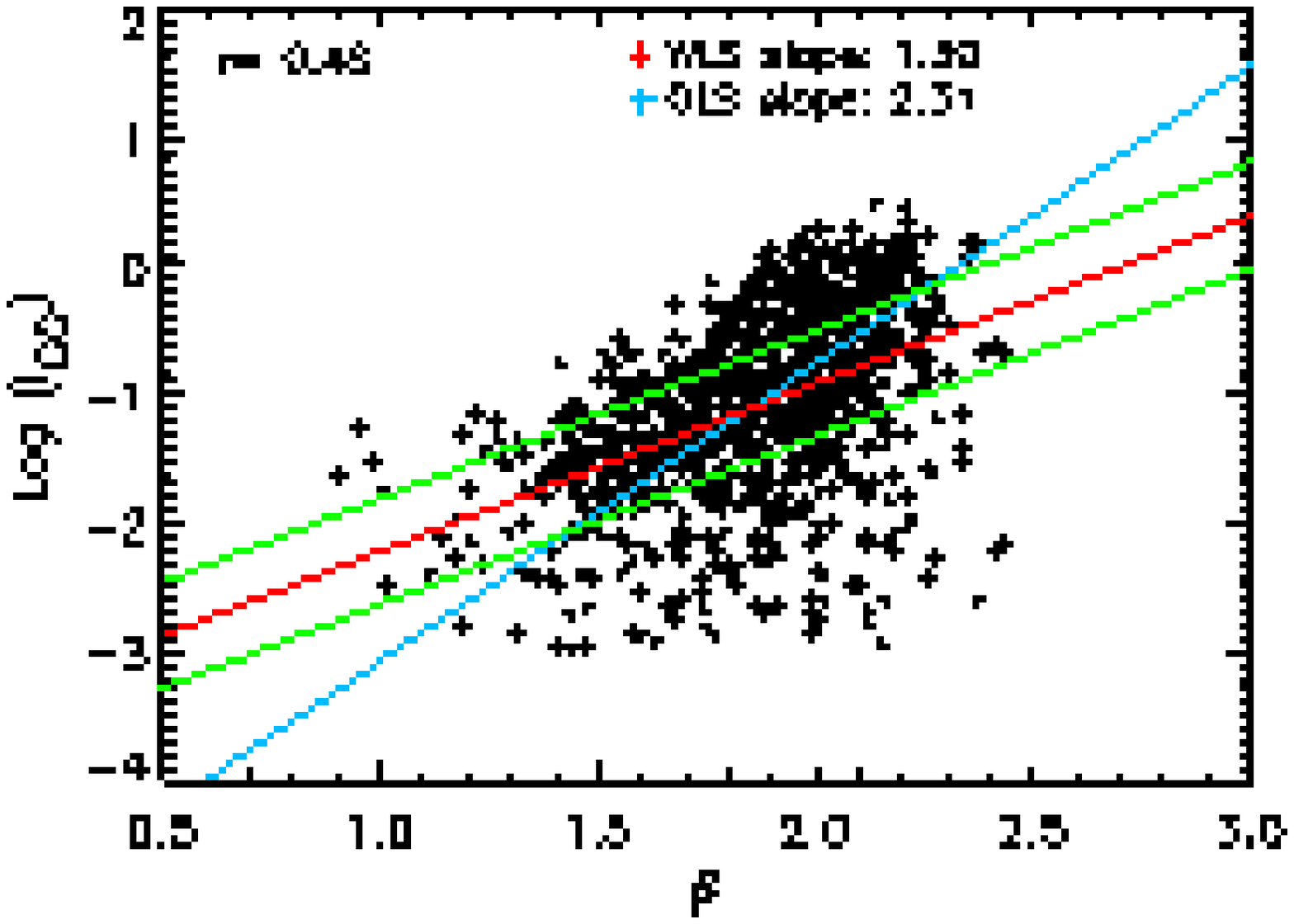}\includegraphics*{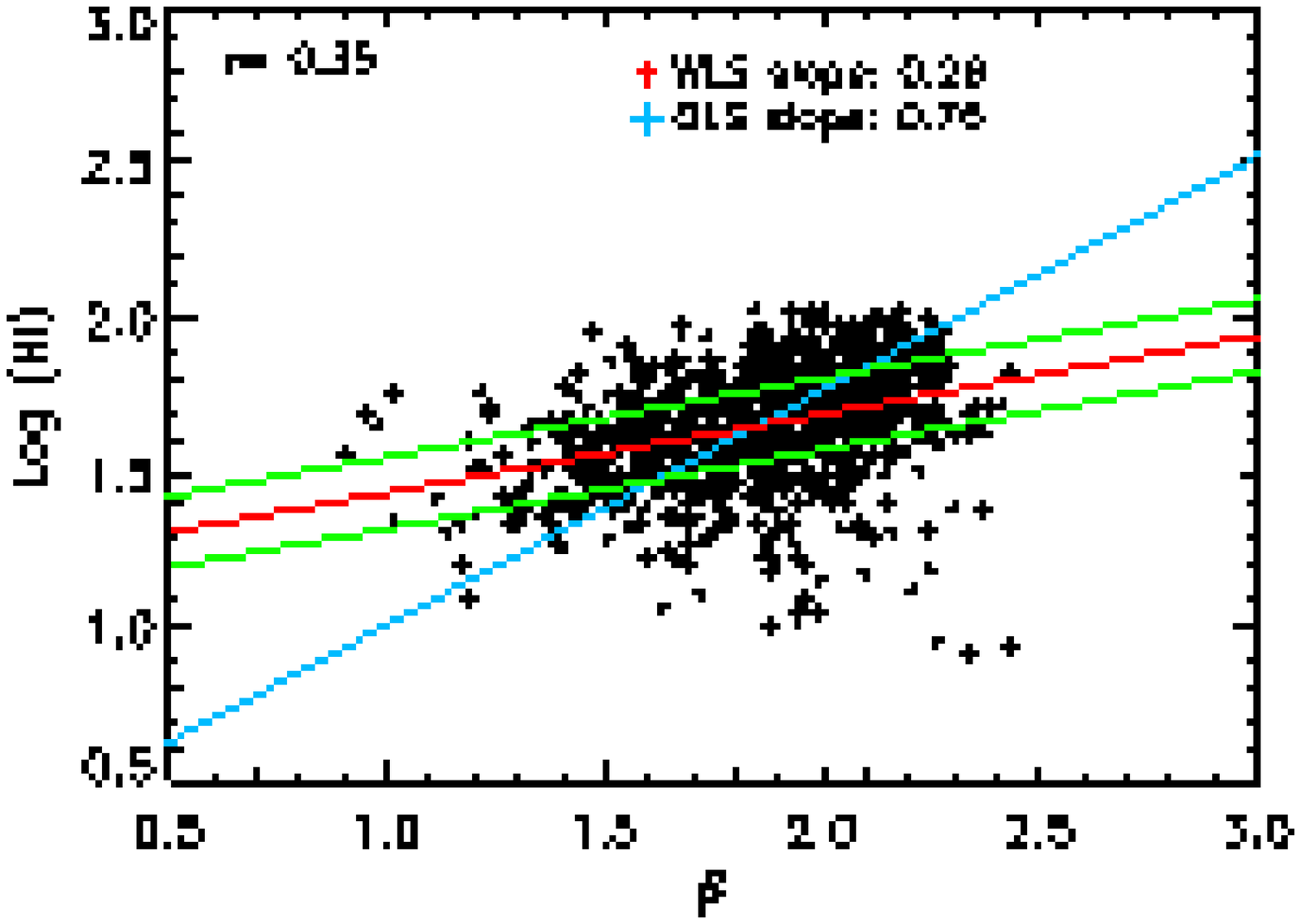}}
\caption[]{Correlations between the dust spectral index and H$\alpha$, CO(2-1), and HI integrated intensities. The WLS weighted least square fit (red line) and its 5-sigma confidence levels (green lines) are shown. The blue lines show the ordinary least square fit (OLS).}
\end{center}
\end{figure}



\end{document}